# topas-editor and TOPilot: tools for AI-driven analysis of diffraction and total scattering data

**John S. O. Evans,[a] Oliver J. Wagstaff,[a] Ivana R. Evans,[a] Arnt Kern,[b] and Alan A. Coelho[c]**

[a] Department of Chemistry, University Science Site, Lower Mountjoy, Durham University, South Road, Durham, DH1 3LE, UK

[b] Bruker AXS SE, Östliche Rheinbrückenstraße 49, 76187 Karlsruhe, Germany

[c] 72 Cedar Street, Wynnum, 4178, Brisbane, Australia

**Synopsis** We describe *topas-editor*, a VS Code editor extension for creating INP files and interacting with *TOPAS*, and *TOPilot*, a set of rules, examples and scripts that lets a large language model drive *TOPAS*. Together they can automatically perform complex crystallographic analyses of powder, single crystal and PDF data, providing INP files and reports for the user to check.

**Abstract** This paper describes *topas-editor* and *TOPilot*, which provide a powerful interface to the *TOPAS* software suite and allow generative AI-driven automation of complex crystallographic tasks. *topas-editor* is a *TOPAS*-aware VS Code extension designed for the rapid creation and editing of *TOPAS* input files and for interacting with the software. It knows *TOPAS* syntax, has crystallographic-aware editing features and provides in-line user guidance via hover tips and automated links to reference manuals. *TOPilot* is a collection of Python routines and machine-accessible reference documents that allow AI-driven automated analysis in *TOPAS*. It uses a transferable *topas-inp-writer* skill that can be applied in all *TOPAS* application areas. We demonstrate the power and reliability of the approach using examples including: a complex Rietveld refinement with sample- and data-related complications; automatic analysis of multiple data sets; a pipeline from indexing through structure solution to refinement; automatic identification of the phases present in a sample; single crystal and powder diffraction analysis of incommensurate structures; and fitting PDF data. *TOPilot* can be used directly alongside *topas-editor*, via a stand-alone *topilot-wizard*, or via command line interfaces or desktop apps from major AI providers. *TOPilot* was developed and tested using Claude large language models, but can be configured to use other cloud or local models. *topas-editor* and *TOPilot* are freely available.

## 1. Introduction

*TOPAS* is a powerful software suite for the structural and microstructural analysis of materials (Coelho *et al.*, 2011; Coelho, 2018; Dinnebier *et al.*, 2018). It is optimised for the analysis of powder diffraction, PDF and single crystal data from X-ray (constant wavelength or energy-dispersive), neutron or electron sources. It can describe both commensurate and incommensurate structures and, with neutron data, both nuclear and magnetic scattering. A variety of methods are implemented to extract microstructural information from powder data. The *TOPAS* minimisation routines are fast, memory-efficient and robust, meaning that complex analyses of thousands of powder patterns with thousands of structural models are possible.

The flexibility of *TOPAS* means that only a fraction (perhaps 5–10%) of its functionality is exposed through its standard GUI. The full functionality is accessed through its INP file format. INP files are plain text files that contain *TOPAS* keywords and their associated arguments, standard or user-defined macros, and user-defined equations or functions to describe any non-standard analysis required. While INP files are human-readable, there is an inevitable learning curve to creating them. For example, *TOPAS* has >700 keywords and >400 bundled macros, and their scope and syntax in the INP file must be correct for *TOPAS* to run. In this paper we describe *topas-editor*, which is an editor designed to create and edit INP files and interact directly with all *TOPAS* versions. We also describe *TOPilot*, which provides a *TOPAS*-aware artificial intelligence (AI) interface for *topas-editor*. *TOPilot* can control the creation of INP files, automatically perform complex crystallographic analyses and archive results. We show how *TOPilot* can, for example, reliably perform complex workflows from indexing through structure solution to full Rietveld refinement with minimal user intervention. *TOPilot* ultimately delivers an INP file of the final model which the user can verify and re-run.

There have been many recent applications of AI to powder diffraction analysis. These have included data-driven machine learning (ML) approaches to indexing and space group determination (Vecsei *et al.*, 2019; Chitturi *et al.*, 2021) and pattern matching and phase identification (Park *et al.*, 2017; Oviedo *et al.*, 2019; Ozaki *et al.*, 2020; Lee *et al.*, 2020; Schuetzke *et al.*, 2021; Dong *et al.*, 2021; Maffettone *et al.*, 2021; Szymanski *et al.*, 2021; Greasley & Hosein, 2023; Szymanski *et al.*, 2024). More recently, various attempts to automate Rietveld refinement (Feng *et al.*, 2019; Ozaki *et al.*, 2020; Szymanski *et al.*, 2023; Biwer *et al.*, 2025; Li *et al.*, 2026; Mun *et al.*, 2026; Shin *et al.*, 2026; Wu & Sun, 2026; Cao *et al.*, 2026; Campbell *et al.*, 2026; Wu & Gaultois, 2026) and more complex workflows (Szymanski *et al.*, 2023; Corrao *et al.*, 2026) have appeared. The importance of careful expert oversight of these approaches has been emphasised (Leeman *et al.*, 2024). In general, data-driven ML methods need a large body of training data, and their ability to generalise depends critically on the scope and representativeness of the training set. In particular, simulated training data may fail to reproduce the combined variability in experimental data introduced by sample purity and preparation, instrumental configurations, microstructure, data processing and measurement conditions, limiting their applicability. General-purpose large language models (LLMs) offer an alternative approach through their capacity for multi-step problem solving and their ability to retrieve information from the literature, to make analysis decisions based on prior findings and to perform repetitive tasks. Coupling these LLM abilities with domain-specific crystallographic software, domain-specific rules and user input when required provides a powerful analysis combination.

*TOPilot* adopts this strategy and mirrors the analysis methods used by an experienced crystallographer: *TOPilot* weighs its refinement decisions against a curated and referenced set of rules (which can be extended by the user), then checks models using *TOPAS*. The speed and robust and reproducible least-squares convergence of *TOPAS* mean *TOPilot* can automatically assess hundreds of different parameter combinations in the structural model to test sensitivity. It makes many analysis decisions using generative AI, but uses hard-coded algorithms for important crystallographic decisions (e.g. symmetry) and will ask the user rather than guess missing information (e.g. experimental details). *TOPilot* can be driven through text or spoken input. The use of AI also means that perhaps 95% of the most common *TOPAS* tasks can alternatively be driven through a simple *topilot-wizard* GUI, where the user has to enter just two or three choices to trigger complex workflows.

The following sections describe *topas-editor* and how the *TOPilot topas-inp-writer* skill was developed and tested. These are followed by brief examples of *TOPilot* performance on real examples and information on how to download and use the packages.

## 2. topas-editor

*topas-editor* is an extension to the free VS Code editor and can be installed directly from the VS Code Marketplace (see Section 7). In a typical setup (Figure 1) a series of *TOPAS*-specific menus appear on the left of the window, the colour-coded INP file with hover help explanations of the syntax appears in the centre, and an optional chat window (see Section 3) appears on the right. The left-hand menus contain options for creating and editing the INP file, sending it to *TOPAS* for analysis, and for configuring the editor. Many common commands have a ctrl-t<letter> shortcut (e.g. ctrl-ta to launch *TOPAS* Academic) and/or are available via a right-click in the INP window. The full list of commands is also available by typing Ctrl+Shift+P followed by "*topas*" in the VS Code Command Palette. Most *topas-editor* commands are written in TypeScript and are installed directly; additional custom user menus are supported. *topas-editor* can also use several of the Python routines that are installed with *TOPilot* without using AI. *topas-editor* features include:

- Full syntax highlighting of *TOPAS* keywords and macros; *TOPAS*-specific folding to simplify editing large files; optional decoration to highlight data sets (xdds) or phases (strs); highlighting of common syntax errors; access to multiple VS Code extensions.
- Interactive tooltips by hovering over keywords either as an in-line summary or via a hyperlink to the full technical manual; additional AI help through *TOPilot.*
- Routines to automatically create symmetry constraints on cell parameters, atomic sites and displacement parameters and set sensible minimum and maximum limits on refined structural parameters.
- A library of INP templates for the most common analysis types; links to hundreds of test examples and automatic searching of examples.
- Insertion of common INP snippets and in-line execution of shell editing commands like grep, sort or sed.
- Menus to create standard INPs (Rietveld, Pawley, etc) through three or four user selections; automatic backup of files.
- Automatic update after INP file execution in *TOPAS*, or zero-click refinement and update via ctrl-tc.

A small subset of these features was available in *jEdit* (Evans, 2010), which the *TOPAS* community used for many years; the *topas-editor* interface is designed to make switching from *jEdit* straightforward.

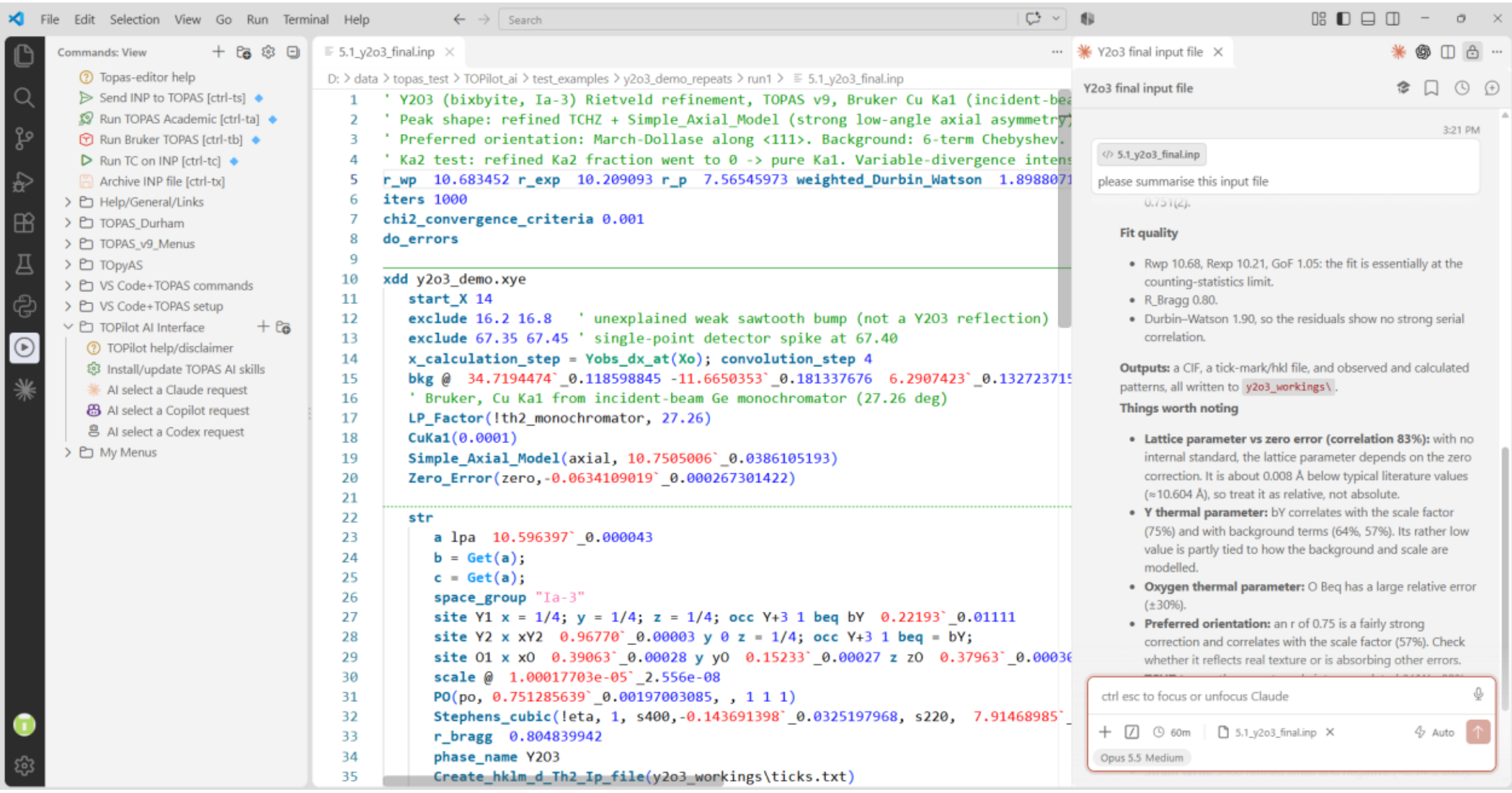

**Figure 1** Screenshot of *topas-editor* window. *TOPAS*-specific menus on the left, INP file in centre and Claude Code agent on the right. The INP file describes a Rietveld fit of $Y_2O_3$ and was auto-generated in Example 5.1. *TOPAS* keywords are bold blue, refined numbers red, fixed numbers or standard uncertainties in blue and comments in green. The Claude Code window contains part of the response to the prompt "*please summarise this input file*".

### 3. *TOPilot* and *topas-inp-writer* setup and development

One reason for building *topas-editor* as a VS Code extension was VS Code's potential for AI integration. GitHub Copilot was bundled with VS Code from late 2024 and native agent extensions for Anthropic (Claude Code) and OpenAI (Codex), which allow LLMs to read files, run software and act on the results, appeared from autumn 2025; Google's Antigravity and other agent extensions also now exist. *TOPilot* uses these external LLMs for intelligent automated analysis in *TOPAS* via the *topas-inp-writer* skill. We use *TOPilot* to encompass the *topas-inp-writer* skill with its scripts and examples, the *topilot-wizard* (Section 4), the *topas-editor* interface and any future *TOPAS* skills. *topas-inp-writer* uses the Anthropic Agent Skills format, which has YAML front matter (a skill name and description as a minimum) followed by plain Markdown (md) text. The main *SKILL.md* file describes *TOPAS* functionality and gives overview instructions on how different types of analysis should be performed. This information is supplemented by more detailed instructions and code distributed in three separate folders: *references*, *scripts* and *example_inp_files*. The skill format is an open standard supported by several agent tools. The bulk of the work reported was done using

Claude Code, which gives access to a variety of Anthropic models, but can also be configured to access other external or cloud LLMs.

The *references* folder contains detailed instructions on different types of *TOPAS* analysis in a set of thirty-plus Markdown files. These contain both the how and why rules *TOPilot* follows. These and the main *SKILL.md* were developed with Claude from sources including: the *TOPAS* Technical Reference manual, a structured listing of the full *TOPAS* keyword hierarchy, the *topas.inc* macros, thousands of example INP files, resources such as the *TOPAS* wiki (https://topas.awh.durham.ac.uk/) and training webpages (https://topas.webspace.durham.ac.uk/tutorial_menu/), and many interactive refinements. Files containing rules and conventions for specific processes (e.g. Rietveld conventions, peak fitting and indexing conventions, etc) were hand-written, then further developed with AI. The *TOPAS* Technical Reference chapters were also reformatted to make them more AI-readable before being condensed into individual md reference files.

One key aspect underpinning *TOPilot's* effectiveness is its understanding of the *TOPAS* data tree that underlies every INP file. Each *TOPAS* keyword creates or sets an item in a hierarchy of objects and items can only appear where the tree allows. For example, a data set (xdd) holds phases (str) that hold sites. The tree is larger than the keyword list, with only about a third of its named items being keywords. A browsable view of the tree was developed that ships with *TOPAS* and tells the model where each keyword may appear and how to interpret names used in equations.

The *scripts* folder mainly contains pre-written Python scripts that are frequently used by *TOPilot* in analyses; bespoke scripts that are generated on-the-fly are stored as scratchpad files for that session. The pre-written scripts cover routine tasks such as plotting results, writing reports, batch analysis of multiple models, checking INP syntax, and tidying up INP files. There are, of course, aspects of crystallographic analysis where decisions should not be delegated to (current) LLMs. One area is crystallographic symmetry, where, for example, a single error in a Wyckoff site makes the model completely incorrect in a way that might not be obvious in an automated Rietveld refinement. The parsing of CIF files and handling of symmetry constraints is therefore always directed through fixed pre-written Python symmetry scripts. The scripts either correct errors and omissions during parsing or provide a specific warning of unresolvable errors. Where necessary, symmetry operators are checked against *TOPAS*'s internal space group generation to ensure consistency of settings and origins.

The *example_inp_files* folder contains curated and annotated INP files for the most common types of analysis. When creating an INP file *TOPilot* checks these first and bases its INP file on them. When no suitable example is found, the wider body of around 300 INP files distributed with *TOPAS* is checked. If there is still no suitable example found, the INP is generated from the rules in the *references* folder.

The *TOPilot* CIF parsing and symmetry constraint routines were tested on a library of over 5000 CIF files from the Crystallography Open Database (COD) (Gražulis *et al.*, 2009), the Cambridge Structural Database (CSD) (Groom *et al.*, 2016), the ICSD (Zagorac *et al.*, 2019) and other sources; COD and CSD CIFs were distributed as evenly as possible across the 230 space groups. *TOPilot* flagged 45 CIFs as containing space group symbols not recognised by the *TOPAS* version used (these have since been added) and around 25 as containing unresolvable errors (hand-checking confirmed this); the rest parse successfully. Automated Rietveld fitting of simulated powder patterns from the 4952 viable CIFs (with and without simulated experimental noise) using *TOPilot* gave perfect fits ($R_{wp}$ close to zero for noise-free data or $\chi^2$ close to 1.0 with artificial noise).

*TOPilot* can be installed and updated from a *topas-editor* menu. The *topas-inp-writer* skill is then invoked whenever a relevant prompt is issued in the Claude Code (or other interface) window. *topas-editor* also contains menus so users can send saved prewritten prompts directly to an LLM without invoking the *topas-inp-writer* skill; this is useful for smaller LLMs and minimising token use. Examples of *TOPilot* analyses and the outputs and reports they create are given in Section 5.

#### 4. *topilot-wizard*

As discussed above, the flexibility and range of applications of *TOPAS* would make a comprehensive traditional GUI too complex to be usable. *TOPilot* ships with a simple GUI which performs many of the most common powder analyses with minimum user input. It can be launched from the scripts folder or via *"/topilot-wizard"* from Claude Code. The user sees a five-box GUI (Figure 2) on a single page, and amber text in the lower left requests any user information required. As a minimum, a single data set or a folder containing multiple data sets is selected. The key experimental information for the powder instrument is then requested. Until further information is entered, the wizard assumes that peak fitting, indexing, structure solution and Rietveld analysis are needed, and prompts the user for an empirical formula. With just this information, "Go" triggers the entire analysis pipeline. If instead a CIF file is selected, or information on a Pawley or peak fitting phase is entered, the peak searching, indexing and structure solution features are disabled and just the whole powder pattern fitting analysis is performed. The pipeline runs either via the VS Code chat window or via the Claude CLI (Command Line Interface). With this minimum input *TOPilot* will make its model choices (e.g. peak shape to use, refinement strategy, etc) according to the rules in the *topas-inp-writer* skill. The user can enter other choices via option menus or in the general text input box. Most of these choices can be saved as defaults for future analyses.

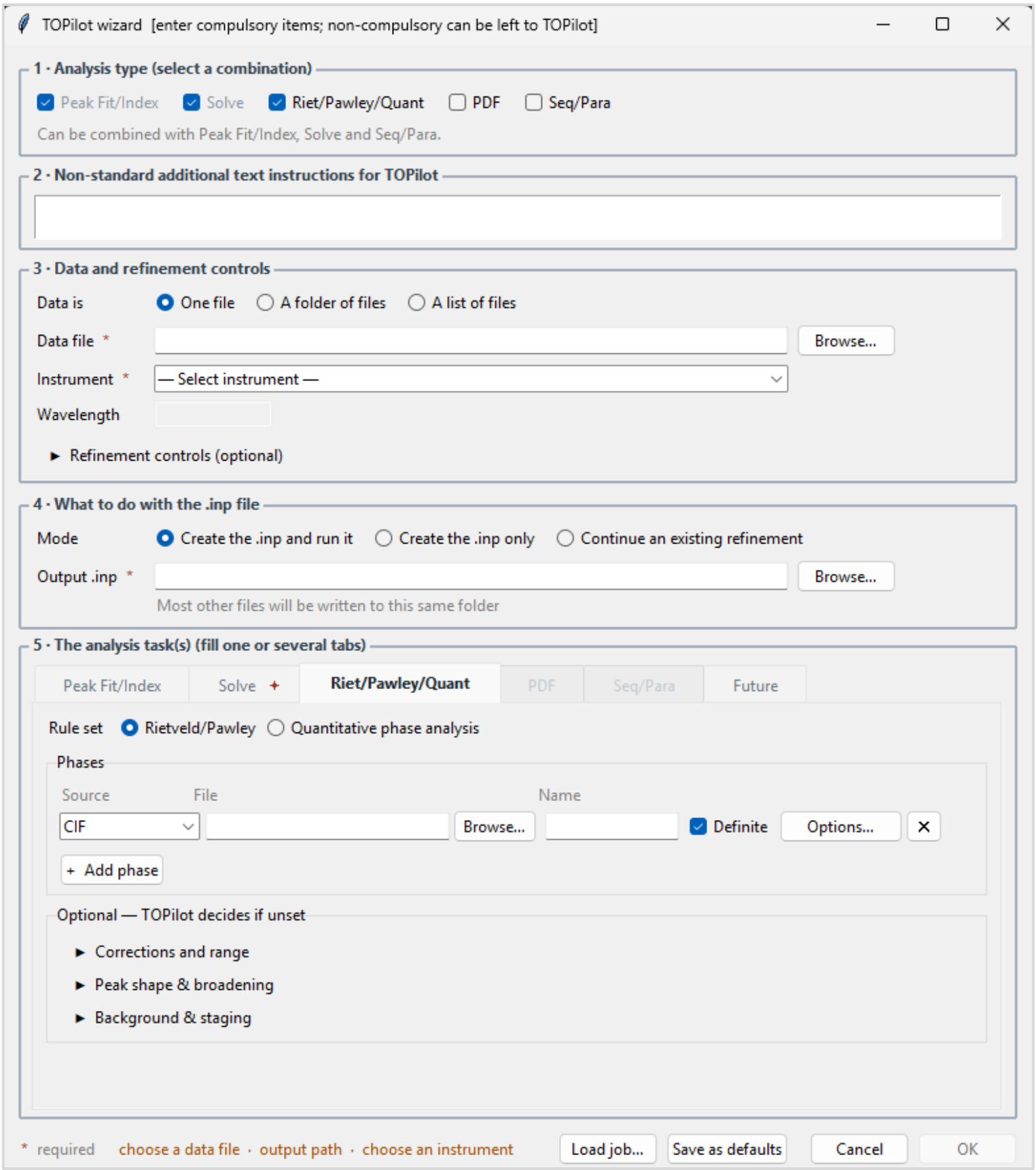

**Figure 2** Screenshot of *topilot-wizard* window. User tasks are in amber in the lower left.

#### 5. Examples

This section describes examples illustrating the use of *TOPilot*. Full prompts used to run each example are provided in the skill *test_examples* folder and SI. In each case *TOPilot* was explicitly instructed not to search for previous examples of the problem, but to rely solely on the *topas-inp-writer* skill. Each example was also run on “clean” computers with a fresh *TOPilot* installation to avoid any influence from stored agent memory. None of the examples described were explicitly used in the *TOPilot* development. Most examples work with *TOPAS*

v7, v8 and v9 (and perhaps earlier versions), though v9 includes features which work more effectively with AI agents and is needed for Example 5.5.

The time and token cost for analyses vary across tasks. For a one-off Rietveld analysis the LLM time dominates and will depend on the model chosen. If thousands of patterns are fitted using a single INP file the LLM time is effectively unchanged and the INP file execution by *TOPAS* (either by the AI or user) will dominate. As an indication, the complex Rietveld analysis in Example 5.1 took around 7.5 minutes with *Opus 5.5* in autumn 2026; the full peak fitting through structure solution and refinement of Example 5.2 took 8 minutes. Both were run on a 5-year-old standard laptop.

#### 5.1. A Rietveld refinement with twists

Data set *y2o3_demo.xye* is a simulated powder pattern of $Y_2O_3$ that we have used at many training schools to illustrate the potential pitfalls of Rietveld analysis to students. The pattern was simulated using a fundamental parameters description for a typical laboratory diffractometer with sample size/microstrain broadening and artificial Gaussian noise added. To challenge students the following aberrations were applied: a 0.2° zero offset, an *hkl*-dependent peak broadening, <111> preferred orientation and a background containing humps and a simulated detector spike. The starting CIF file also has an incorrect cell parameter and metal coordinates shifted by ~0.2 Å from their true values. *TOPilot* was run by telling it to *"perform a Rietveld fit from starting model y2o3_start.cif checking for preferred orientation, hkl-dependent peak shapes and artefacts in the data"* (this and other verbatim prompts are in SI). Despite the example being set up such that a "default" Rietveld approach would diverge, *TOPilot* solved the problem without any intervention, automatically producing the interactive plots shown in Figure 3 and a detailed report of the analysis. The strategy followed included starting with a narrow $2\theta$ range and comparing observed and calculated peak positions ahead of refinement, helping identify the $2\theta$ offset required. After convergence over a narrow range, the $2\theta$ range was extended before peak shape anisotropy and preferred orientation were checked. Identification of "W" or "M" shaped features in the difference profile was used to confirm the peak shape anisotropy and a series of tests performed with different spherical harmonic corrections to better fit peak shapes and intensities. After determining the preferred orientation spherical harmonic coefficients needed to fit the data *TOPilot* chose to test a <111> March–Dollase correction, which uses fewer parameters, and concluded that the simpler model (which was actually used in the simulation) was sufficient. The detector spike was automatically identified and excluded from the analysis. Finally, potentially correlated parameters (identified from the variance–covariance matrix) were reset and re-refined to ensure the global minimum was found, then the final Rietveld fit was compared to a Pawley fit to the same data. The final model and fit quality equal a by-hand analysis.

Reproducibility was checked by running the same analysis a further five times using *Opus 5.5* (medium effort was used unless stated). Four of the five refinements used an identical $2\theta$ range and gave essentially identical $R_{wp}$ and $\chi^2$ values (10.674 to 10.683% and $\chi^2$ from 1.045 to 1.046); the fifth started the fit from a higher $2\theta$ (18 *vs* 14°) giving a slightly lower $R_{wp}$. All refinements successfully identified the preferred orientation and peak shape anisotropy and gave equivalent final structural models. The only other model variations were slight differences in the peak shape parameter set refined and three refinements applying a $2\theta$ zero-point correction while two used a sample displacement correction. These are all parameters that are known to be highly correlated, and which are normally chosen by the crystallographer. In standard *TOPilot* work they would usually be specified in the analysis prompt. The smaller *Sonnet 5.5* model achieved equivalent results with the same minor variability in non-critical parameter choices over five runs. Overall *Sonnet* analysis times were around 33% shorter (5 minutes per run) and approximately 50% cheaper in terms of token use.

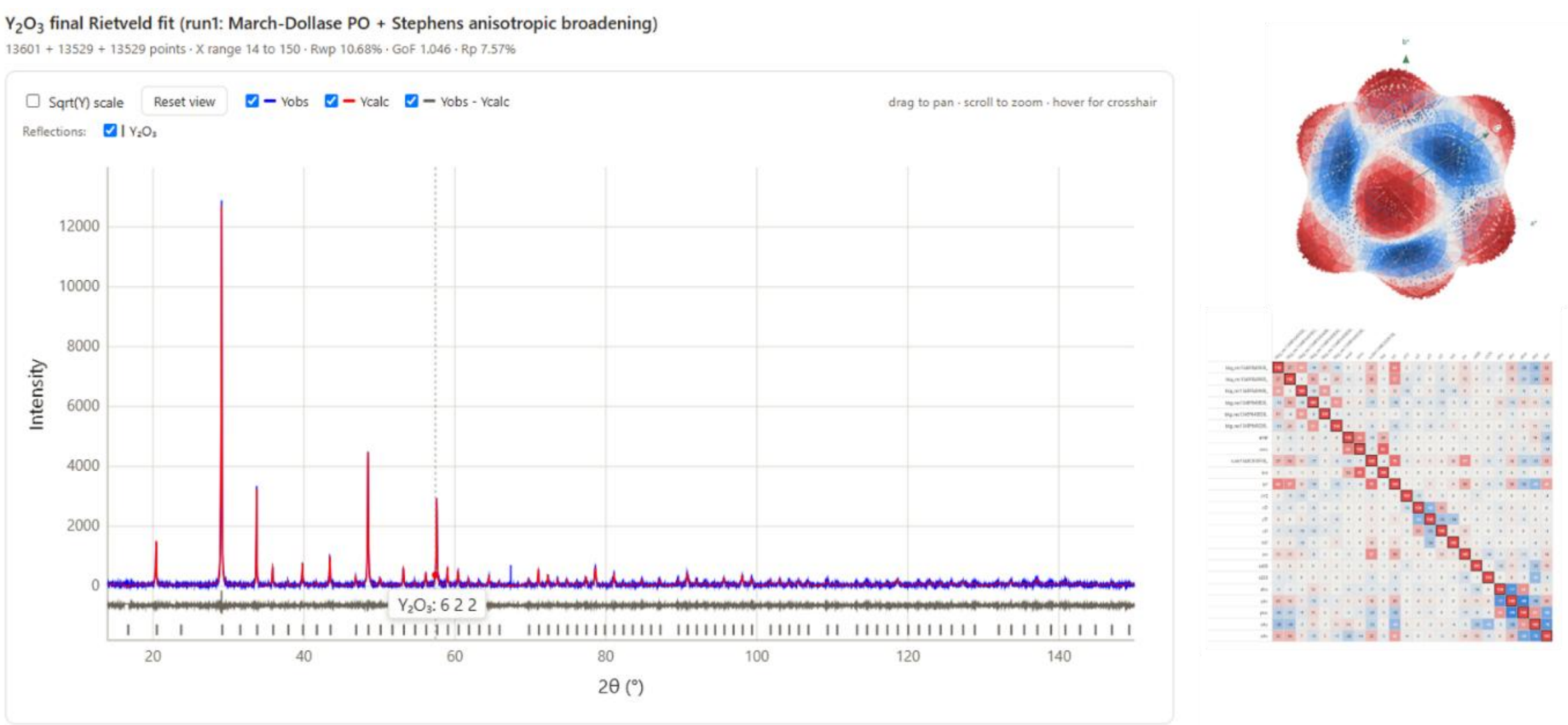

**Figure 3** Automatically generated interactive HTML Rietveld plot, representation of the preferred orientation spherical harmonic tested and variance–covariance matrix heatmap for the $Y_2O_3$ Example 5.1.

#### 5.2. Peak fitting, indexing, structure solution and Rietveld refinement

Example 5.2 (*zrw2o8_index_solve_refine*) illustrates the pipeline from peak fitting through indexing and space group determination to Rietveld refinement using a fixed wavelength neutron powder diffraction pattern of $ZrW_2O_8$ (Mary *et al.*, 1996). If it is executed via the *topilot-wizard* the input required is: (a) selection of the *zrwneut_sh.xy* data set and wavelength (1.8857 Å); and (b) input of the $ZrW_2O_8$ empirical formula. Clicking "Go" then initiates the full workflow. The correct unit cell (cubic 9.16 Å) and space group ($P2_13$) are identified and structure solution via simulated annealing and refinement proceeds smoothly. The final Rietveld fit ($R_{wp}$ 9.06%) was automatically compared with a Pawley fit ($R_{wp}$ 8.95%) over the same $2\theta$ range. The neutron model can also be used to fit X-ray powder data. Here *TOPilot* can be told to "*set up bond distance & angle restraints*" and will automatically restrain the $ZrO_6$ octahedra and $WO_4$ tetrahedra to sensible geometries. An equivalent workflow can be run on several of the other test examples (e.g. $TiO_2$, $LaMnO_3$ X-ray examples).

### 5.3. Sequential Rietveld refinement

Example 5.3 (*variable_temperature*) analyses a series of thirty rapid powder patterns of an $Al_2O_3$/Si mixture heated from 30 °C to 950 °C which was used for furnace temperature calibration on the Diamond I11 beamline. *TOPilot* was provided with the experiment logbook containing filenames and temperatures, the synchrotron wavelength and was asked to save the analysis in a format suitable for plotting at a later date in Reel (Gjørup *et al.*, 2021). *TOPilot* started the analysis with a single overall peak shape, then switched to one peak shape per phase to account for the additional sample broadening of $Al_2O_3$. After establishing an appropriate model for fitting the 30 °C data, a sequential INP file was automatically generated using the *TOPAS* #list syntax to fit all the data and output key parameters. The HTML file showing thermal expansion against set temperature shown in Figure 4 was produced automatically.

The 30 °C data can also be used to explore the *TOPilot* indexing procedures. An automatic peak search, peak fit then examination of FWHM vs $2\theta$ separates the observed reflections into two likely sets. *TOPilot* is then able to index the two peak sets automatically, giving correct unit-cell parameters for both phases.

**$Al_2O_3$ + Si variable-temperature series**

Sequential Rietveld refinement of 30 synchrotron patterns, λ = 0.8257653 Å, 30–950 °C (logbook temperatures).

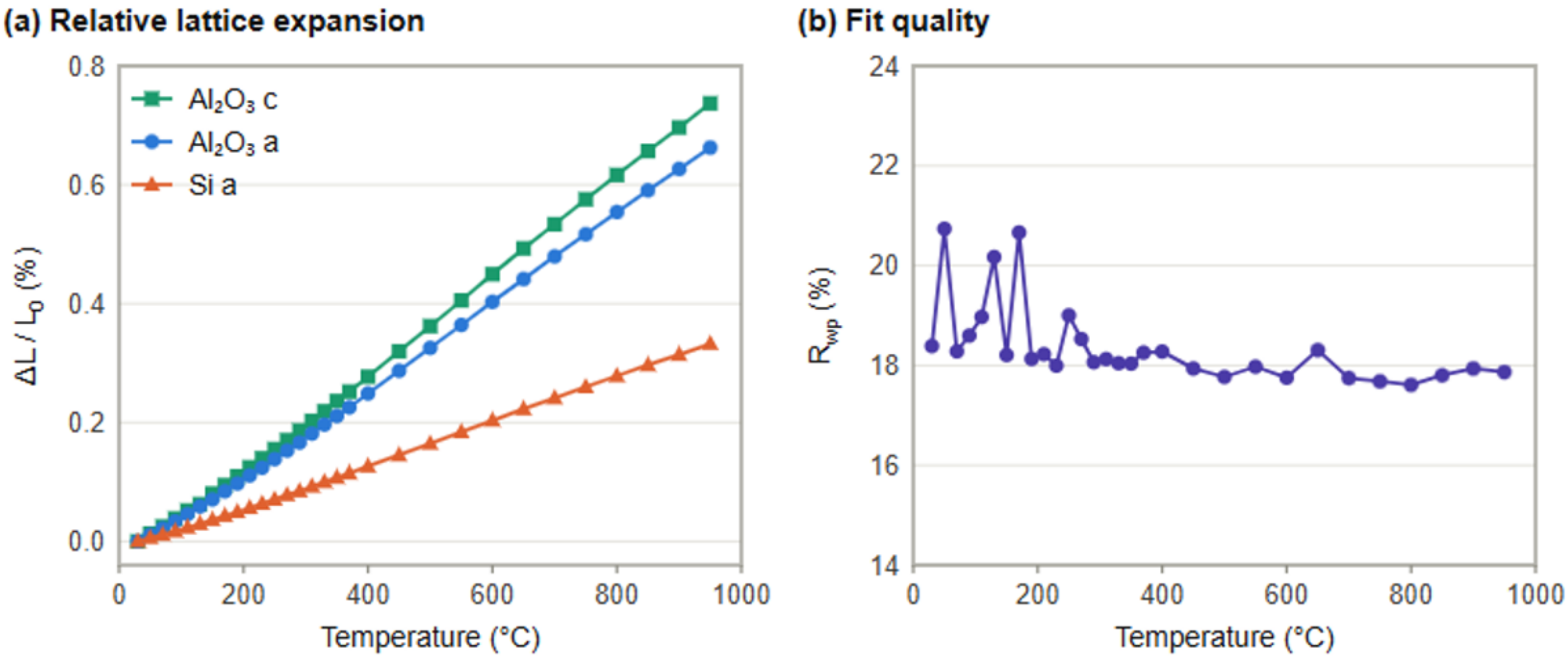

$\Delta L/L_0$ is relative to each axis at 30 °C. Lattice-parameter esds (~$1\times10^{-5}$ Å) are much smaller than the markers.

▼ Data table (refined values, esds in parentheses)

| T (°C) | $R_{wp}$ (%) | GoF | $Al_2O_3$ *a* (Å) | $Al_2O_3$ *c* (Å) | $Al_2O_3$ *V* (Å$^3$) | Si *a* (Å) | Si *V* (Å$^3$) |
|---|---|---|---|---|---|---|---|
| 30 | 18.39 | 1.50 | 4.759248(9) | 12.99223(4) | 254.8537(11) | 5.430966(6) | 160.1885(6) |
| 50 | 20.74 | 1.68 | 4.759767(10) | 12.99393(4) | 254.9427(12) | 5.431286(7) | 160.2168(6) |
| 70 | 18.28 | 1.49 | 4.760284(9) | 12.99548(4) | 255.0284(11) | 5.431570(7) | 160.2419(6) |
| 90 | 18.60 | 1.51 | 4.760849(9) | 12.99724(4) | 255.1236(11) | 5.431899(7) | 160.2711(6) |

etc.....

**Figure 4** Example 5.3 automatically generated interactive plots summarising the analysis. The screenshot above was generated by the additional prompt "*produce a summary plot of key results that could be included in a report*".

### 5.4. Phase ID and quantitative analysis

*TOPilot* has a set of rules to identify phases present in a powder pattern via Rietveld refinement and/or to perform quantitative analysis. The *pigment_example* automatically analyses three powder patterns collected on a lab diffractometer (routine ~25 minute scans) on samples believed to contain one or more of the ancient pigments: Han Blue, Egyptian Blue or Han Purple. The diffractometer logbook recorded the sample ID and data filename and included the CIF filenames for phases possibly present in each sample. Automatic analysis correctly identified that two samples contained essentially pure Han Blue or Egyptian Blue (with strong texture), while the third was a mixture of Han Blue and Han Purple. In the two-pigment sample, a series of unfitted peaks were automatically identified as quartz. *TOPilot* checked the chemical plausibility of this against a web search of the literature, then included quartz as an additional phase. Essentially identical results were achieved over five repeat analyses. With this approach *TOPilot* can automatically analyse data sets as they are recorded. Similar protocols can be followed to analyse (for example) the quantitative round robin data of Madsen & Scarlett (example *quantitative_analysis*) (Scarlett *et al.*, 2002).

### 5.5. Incommensurate structures

*TOPAS* v9 has the capability to analyse modulated structures, with automatic creation of modulation amplitude symmetry constraints. The *topas-inp-writer* skill has a reference document describing this capability and can create modulated INP files automatically. With the *na2co3_modulated* example, *TOPilot* was given the prompt: "*$Na_2CO_3$.cif contains the published modulated structure of γ-$Na_2CO_3$ and its observed reflections. Set up a TOPAS v9 single crystal refinement of that model against those reflections. Hold the published amplitudes fixed at first and refine the scale and extinction; then refine the amplitudes with add_constraints, and report R(F) per satellite order against the published values.*" The γ-$Na_2CO_3$ (Dušek *et al.*, 2003) CIF file supplied was downloaded from the Bilbao Incommensurate Structures Database (Aroyo *et al.*, 2006; Gabirondo-López *et al.*, 2024) and contains the (3+1)-dimensional model in superspace group $C2/m(\alpha 0 \gamma)0s$, with displacive waves to the fourth harmonic and modulated displacement parameters, and 14118 single crystal reflections with satellites to sixth order.

From the CIF, *TOPilot* created an INP file that uses a single model to fit one *hkl* file per satellite order and imported the superspace operators from the CIF (using keyword *mod_tau_from_cif*). With only the scale and extinction refined (published amplitudes fixed), the structure factors agreed with those published to better than 1% in every satellite order up to the fourth. $R_F$ was 3.80, 4.29, 6.87, 10.92, 19.12% for satellite orders 0 to 4, and 6.17% overall; the published model gives 3.81, 4.13, 6.85, 10.91, 19.25%, and 6.15% overall. *TOPilot* then introduced the *add_constraints* keyword to instruct *TOPAS* to refine modulation amplitudes as allowed by symmetry. *TOPAS* assigned 144 free amplitudes (the same as the published model), and refined values were within 2.6 standard uncertainties of those published. The overall process initiated seventeen separate *TOPAS* runs.

The *incommensurate_powder* example uses powder X-ray and neutron data from the incommensurate material $Ce_2O_2MnSe_2$ (Wang *et al.*, 2015), to illustrate the complexity of powder diffraction analysis possible with *TOPilot*. The example was first run before *TOPilot* had been given any instructions beyond those in the reference manual on how to analyse modulated structures, and the prompts used inadvertently failed to provide some of the information needed for the analysis. This gives an indication of how *TOPilot* performs with incomplete information. In a first stage *TOPilot* was told the unit-cell parameters of the parent tetragonal structure and asked to "*find the modulation vector from any weak satellite peaks in the Cu Ka2 X-ray pattern d7_06306.raw recorded with variable divergence slits. Try simple directions first.*". After an automatic Pawley then peak fitting process, the correct q = (0.08, 0.08, 0.5) modulation vector was identified as explaining all the sharp weak peaks in the pattern except a single peak identified as the Cu $K_\beta$ component of the strongest reflection. More general vectors were tested, but were not necessary to fit the data.

In stage two, the parent structure and modulation vector were used to create a CIF file of the modulated structure using *ISODISTORT* (Campbell *et al.*, 2006). This was done by hand but could be automated. *TOPilot* was then asked to perform a combined Rietveld fit of the X-ray pattern and a neutron powder diffraction pattern recorded on the GEM diffractometer (Williams *et al.*, 1997). It was given the GEM calibration constants and the instrumental peak shape description. It was asked to fit first order satellites using a Crenel function to describe the occupational modulation of Mn. We made four accidental errors or omissions in the prompt: (1) the X-ray diffractometer monochromator was not specified (it was not needed in stage one of the analysis); (2) the prompt on the occupational modulation was unclear; (3) a typo was made in the macro name for the instrumental peak shape; (4) we forgot to specify the correct weighting scheme for these neutron data, which have a very different $y_{obs}$ range to the X-ray data (X-ray $y_{obs,max}$ = 160,000; neutron $y_{obs,max}$ = 2.9). *TOPilot* prompted the user for further input on points (1) and (2), and automatically identified the correct macro to use (informing the user) for point (3). It then automatically set up and performed the refinement, making correct choices for the symmetry-allowed modulation amplitudes. The vastly different contributions to the overall $\chi^2$ from the X-ray and neutron data were automatically identified and *TOPilot* chose to repeat the refinement applying Poisson statistics for the neutron data (again, warning the user). The final fit is identical to that performed independently by hand, converging to $R_{wp}$ = 2.96%, and took around 15 minutes (*Opus 5.5*). A significant portion of that time was spent testing different parts of the Rietveld model that could have been excluded through the prompt. For example, *TOPilot* tested several models for anisotropic parent peak shapes and for additional satellite broadening before concluding these were not statistically significant. A well-structured INP file, reports and interactive Rietveld fits were automatically produced. From this point the fit could be further developed either interactively through *TOPilot* or by hand in *topas-editor*.

### 5.6. PDF fitting

Example 5.6 (*pdf_data*) exemplifies PDF fitting using a simulated neutron PDF data set from the *TOPAS* online tutorials (https://topas.webspace.durham.ac.uk/zrw2o8_pdf_adps/). The prompt issued was: "*PDF data are in zrw2o8_topas_sim_gr.xy and a starting structural model in zrw2o8_start.cif. Fit data from 0.5 to 50 Angstroms but apply a weight of zero below 1.5 Angstrom*s" as per the tutorial instructions. In this prompt we inadvertently forgot to specify that the data are a neutron simulation. In its first analysis attempt on PDF data *TOPilot* automatically identified this error from the high $R_{wp}$ obtained, switched to using neutron scattering lengths and informed the user. The final $R_{wp}$ of 2.5% is the best achievable with these data and a standard anisotropic model.

## 6. Discussion and conclusions

This paper describes a powerful and flexible environment for applying generative AI to analysing diffraction (powder and single crystal) and total scattering data. The approach combines the general reasoning and workflow-generation capabilities of modern large language models with the rigorous crystallographic modelling and refinement capabilities provided by *TOPAS*. *TOPilot* can currently tackle all *TOPAS* application areas (peak fitting, indexing, structure solution, whole powder pattern fitting, different data types, different microstructural models, rigid bodies and z-matrices, etc) and can be given additional problem-specific instructions as needed. It has been successful for the examples highlighted in Section 5, all the problems on the *TOPAS* tutorials website (https://topas.webspace.durham.ac.uk/topas_user_menu/) that we have tested and many others. As with any AI tool, its outputs (analysis reports and a working INP file) should be treated as a suggestion for the user to check.

Our main conclusions are:

(1) The *TOPilot* routines as installed and (optionally) executed via *topas-editor* correctly perform many types of complex crystallographic analysis with minimum user input.

(2) *TOPilot* is deliberately built to use verified scripts for parts of the analysis (e.g. symmetry constraints) to avoid AI errors and to refer back to the user on any decisions that should not be guessed (e.g. experimental details).

(3) Reproducibility is excellent on all the examples we have tested. Any minor differences between repeat runs are similar to the different choices experienced crystallographers would make in the absence of further information (e.g. choice of a zero point or sample height displacement; which of several correlated parameters in a peak shape description to fix).

(4) The *TOPilot* / Claude Code (or other agents) approach through *topas-editor* gives a unified environment for both AI and interactive work. While the AI is working the user can review and test intermediate INP files.

(5) *TOPilot* can be a valuable teaching aid to explain the syntax or scientific intent of INP files.

(6) AI tools excel at testing multiple refinement models, using a similar approach to code debugging and testing. Combining this with the rapid and reproducible convergence of *TOPAS* means that a large parameter space of Rietveld models can be efficiently and reliably explored to find the global minimum. There is an old adage in Rietveld refinement that a refinement is never finished – you just give up trying models. AI's perseverance and willingness to consult manuals often exceed those of normal users!

(7) *TOPilot* provides detailed summaries of each refinement and writes interactive HTML files so that fits can be examined at a later date without use of *TOPAS*. It warns the user about the limitations of the model and explicitly flags any major assumptions that require expert assessment.

(8) *TOPilot* can be easily extended to other analysis types or adapted by users to their own preferences by editing or creating simple text files. *TOPilot* can also use existing keywords and macros to generate new user-defined refinement models.

(9) The final responsibility for any analysis must rest with the user, but even in situations where *TOPilot* makes different initial choices to a user's preference, the increase in productivity via automated INP generation and testing is significant. Model changes typically require only minor edits of INP files.

(10) In addition to user-verification of *TOPilot*-generated INP files, we strongly recommend *TOPilot* checks of all user-generated INP files.

**7. Software availability and usage**

*topas-editor* (current version 2.0.3) can be installed directly from the VS Code Marketplace or via https://topas.webspace.durham.ac.uk/vscode_setup/. *TOPilot* (current version 1.1.3) can be downloaded from GitHub or more conveniently via the "Install/update TOPilot AI skills" menu item in *topas-editor*. *topas-editor* can be set to auto-update with new releases, but *TOPilot* updates are currently triggered manually to prevent accidental overwrite of information in the user's ~*.claude\skills\topas-inp-writer* directory. *TOPilot* development was mainly done using Anthropic *Sonnet* or *Opus* 5 models. At the time of writing the token usage required to run analyses using *Sonnet 5* is such that around 80% of a five-hour window's token allocation of a Claude Pro account (about US$20 per month) is typically used in five hours of typical analysis.

### 8. Use of AI in writing the paper

This paper was written by combining cutting-edge technologies from the 19$^{th}$ and 21$^{st}$ centuries. The first draft was written with a fountain pen and paper then scanned. AI was used to digitise the document, format it to the journal template, check for typographical errors and merge comments then tracked changes from authors.

**Acknowledgements** We thank Phil Chater, Dean Keeble and James Gilbert for early discussions around VS Code; and Fábio Furlan Ferreira for sharing his *topas-editor* scripts.

**Conflicts of interest** A.A.C. develops *TOPAS* and *TOPAS*-Academic and receives income from them. A.K. works for Bruker AXS SE, who sell the commercial version of *TOPAS*. The other authors declare no conflicts of interest.

**Data availability** All the examples cited are available as part of the standard *topas-inp-writer* distribution in its *test_examples* folder (GitHub: https://github.com/johnsoevans/topas-inp-writer).

# Supporting information

### S1. Data and prompts

All test examples cited are included with the *TOPilot* distribution. Specific prompts used to run the examples as described in the text are given below. These are typically more verbose than required as they are written as teaching examples. Some prompts are more general than would be used in practice to illustrate the decision making process followed by *TOPilot*. Typographical errors in the prompts have not been corrected.

Each prompt was preceded by instructions to minimise any agent memory of the problems or use of extensive information outside the skill:

1. Forget any previous work you've done on this example and start from scratch
2. Use the minimum number of prompts this session, allow anything you can.
3. Don't look at other files on this computer which may have previous examples of the problem.
4. Don't look on the internet for specific earlier examples of this problem.

#### S1.1. Example 5.1

1. The data is in y2o3_demo.xye in the same directory as this md file.
2. The cif file with a starting model is in y2o3_start.cif.
3. Data were recorded on a Bruker diffractometer with Cu Ka1 radiation and a Ge monochromator.
4. Store any files apart from the inp file in a folder y2o3_workings.
5. Consider the background treatment carefully.
6. Check for preferred orientation.
7. Check for hkl-dependent peak shapes.
8. Check for any instrumental artifacts.
9. Check carefully that you have followed all these instructions and all those in the Rietveld skill.

#### S1.2. Example 5.2

Information supplied via the *topilot-wizard*: Data file zrwneut_sh.xy. Neutron CW data. Wavelength 1.8857 Å. Empirical formula Zr W2 O8.

Follow up: Can you now perform a Rietveld refinement starting with this model using data set 6G17PB01.raw which was recorded with Cu Ka2 radiation, no monochromator and needs a variable divergence slit intensity correction? Use a TCHz peak shape. Add restraints on the ZrO6 octahedra and WO4 tetrahedra.

#### S1.3. Example 5.3

1. This is a variable temperature synchrotron experiment recorded on a Si and $Al_2O_3$ mixture as a temperature calibrant.
2. It was synchrotron data with a wavelength of 0.8257653 Anstroms.
3. The cif files are in this directory.
4. The data files are in a folder call xye_files.  I downloaded them from https://topas.webspace.durham.ac.uk/tutorial_multitopas/.
5. The logbook containing the filenames and temperatures (in C) is below.
6. Please analyse them all and plot the cell parameters of Si and $Al_2O_3$ as a function of temperature.
7. I don't need any quantitative phase analysis.
8. I'm going to want to look at the refinements afterwards in Reel.
9. Save the main input file and results in this folder.  Open the input file when it's made.
10. Store any temporary files in al2o3_si_workings folder.
11. I will want to run the file myself in TOPAS after you're done.

#### S1.4. Example 5.4

1. This is an automatic pipeline for checking the id of phases present in datafiles measured on the diffractometer.
2. Use Rietveld analysis but there is no need to do a structure refinement unless you think absolutely necessary (e.g. no coordinate refinement needed).
3. Please check if each pattern contains the expected phase.  Suggest any issues across the series.
4. Describe the sample peak shape contribution using size/microstrain broadening. There is no need to test hkl-dependent broadening unless absolutely necessary.
5. The data filenames and cif files of structures possibly present are in the table above from the diffractometer logbook.
6. Data were collected on a diffractomer with Cu Ka1/Ka2 radiation and no monochromator. Variable divergence slits were used.  The instrument resolution function can be described with: Simple_Axial_Model(!axial, 7.2) TCHZ_Peak_Type(!pku,-0.000363,!pkv, 0.00129,!pkw, -0.001325,!pkz, 0.0000,!pky, 0.02932,!pkx, 0.01351).

7. Create the main inp files in this directory and open as soon as possible.
8. Store other files in a folder pigment_workings.
9. In the final report include the time taken for the analysis, the number of topas runs and the total number of tokens used.

#### S1.5. Example 5.5

Single crystal example:

$Na_2CO_3$.cif contains the published modulated structure of $\gamma$-$Na_2CO_3$ and its observed reflections. Set up a TOPAS v9 single crystal refinement of that model against those reflections. Hold the published amplitudes fixed at first and refine the scale and extinction; then refine the amplitudes with add_constraints, and report R(F) per satellite order against the published values.

Powder data modulation vector search:

1. d7_06306.raw was recorded on an incommensurate material using Cu Ka2 rediation and variable divergence slits.
2. The strong peaks can all be described with a tetragonal unit cell with space group P4/nmm and cell parameters of a = 4.01 and c = 9.11 Angstroms.
3. There is a known impurity peak at around 28.75 degrees 2-theta. I suggest excluding 28.5 to 29.0 from the fit. There is significant background scatter below 8.5 degrees so I suggest using 8.5 to 50 degrees for the analysis.
4. Can you determine the modulation vector from the weak satellite peaks? Try simple directions first then more complex ones.

Combined fit to X-ray and neutron data:

1. Based on this modulation vector I have used ISODISTORT to produce the cif file cefe0.5ose_01_q_0.08_0.08_0.5_s4_Cmma(a,0,0.5)0s0.cif.
2. I would like to try a combined Rietveld refinement against the X-ray data and a neutron powder diffraction pattern. Only first order satellite reflections are observed.
3. The neutron data were collected on the GEM diffractometer.
4. The GEM instrument scientist suggested the following peak shape parameters and instrumental constants:

wifd_mic_new2(two_theta, =bank_two_theta;, !tspow, 2.75914,!tauf_0, 1.38544, !tauf_1, 1.05054, !taus_0, 0.02609, !taus_1, 0.01029, !t_eff, 233.59810, !hhh, 0.00257, !double_pulse, 0.01015)

TOF_x_axis_calibration(!t0_bank4,-2.6912, !difc_bank4, 4835.91, !difa_bank4,-0.7273`)

local !bank_two_theta 63.62

5. Refine modulation amplitudes as allowed by symmetry.
6. Use a Legendre approach for Mn occupancy modulation with a crenel center of 0.75 and width of 0.5.

**S1.6. Example 5.6**

1. (Neutron) PDF data are in zrw2o8_topas_sim_gr.xy and a starting structural model in zrw2o8_start.cif.
2. Fit data from 0.5 to 50 Angstroms but apply a weight of zero below 1.5 Angstroms.